\documentclass[aps,prb,preprint,groupedaddress,floatfix]{revtex4}
\usepackage{latexsym}
\usepackage{amsmath,amssymb,amsfonts,textcomp}
\usepackage{graphicx}
\usepackage{natbib}
\usepackage[T1]{fontenc}

\def\m#1{\mathrm{#1}}
\def\diam{{\mit\Phi_{\mbox{\footnotesize{\sc{id}}}}}}
\newcommand{\greeksym}[1]{{\usefont{U}{psy}{m}{n}#1}}

\begin{document}
\title{Low frequency acoustics in solid  {\raisebox{0.8 ex}{\normalsize{\bf
        4}}}He at low temperature } 
\author{Yu.~Mukharsky, A.~Penzev, and  E.~Varoquaux}
\affiliation{CEA-DRECAM, Service de Physique de l'\'Etat Condens\'e, \\
  Centre d'\'Etudes de Saclay, 91191 Gif-sur-Yvette Cedex (France)}
\date{\today}

\begin{abstract}
  The elastic properties of hcp $^4$He samples have been investigated using
  low frequency (20 Hz to 20 kHz) high sensitivity sound transducers. In
  agreement with the findings of other workers, most samples studied grew very
  significantly stiffer at low temperature; Poisson's ratio $\nu$ was observed
  to increase from $0.28$ below 20 mK to $\sim 0.35$ at 0.7 K.  The span of the
  variation of $\nu$ varies from sample to sample according to their thermal
  and mechanical history. Crystals carefully grown at the melting curve show a
  different behavior, the change in $\nu$ taking place at lower $T$ and being
  more abrupt.
\end{abstract}
\maketitle

The prospect that solid $^4$He may exhibit a new state of matter,
the supersolid state, with both mechanical rigidity and superflow behavior, is
arousing strong renewed interest in its mechanical properties below 0.2 Kelvin
\cite{reviews}.  The supersolidity features are seen primarily in torsional
oscillator (TO) measurements of the moment of inertia, which decreases below
$\sim$ 0.2 K or so in a way that can not be accounted for in the framework of
classical mechanics, the so-called ``non-classical rotational inertia'' (NCRI)
effect \cite{Leggett:70}. 

In this work, we are concerned primarily with the elastic properties of hcp
$^4$He below 0.8~K.  The velocities and attenuation of sound in solid $^4$He
have long been known to display anomalies at low temperature. Early studies
\cite{Wanner:76,Tsymbalenko:77,Iwasa:79,Tsuruoka:79,Paalanen:81,Hiki:83} have
attributed these anomalies to the presence of dislocation lines, even in good
quality crystals, with typical line densities of $\Lambda \simeq10^6\; \m{
  cm}^{-2}$ and internodal lengths of the dislocation network $L_{\m N} \simeq
5\; 10^{-4}$ cm. Already in these early studies, it was noted that sample
preparation greatly influences these properties \cite{Iwasa:79} and could lead
to anomalous values of $\Lambda$. Nonetheless, these anomalies have been
interpreted in the framework of the usual theory of
elasticity~\cite{Paalanen:81,Hiki:83}; they find a natural explanation in
terms of dislocation motion in the framework of the Granato-L\"ucke theory, as
discussed by many authors over the years (see, e.g.
\cite{Day:09,Beamish:85,Syshchenko:09} and references therein)

The more recent work was directed toward the search of clear-cut quantum
effects and, possibly, to a direct manifestation of supersolidity
\cite{Goodkind:02}, which is now thought to be found in TO measurements
\cite{Kim:04}.  The anomalies in the shear compressibility are possibly
related to those in TO experiments, a possible linkage that makes the subject
matter of an ongoing debate \cite{Clark:08,Day:09}. Here we probe directly the
low frequency elastic properties of solid $^4$He at unprecedented low levels
of strain in the solid ($\epsilon_{zz} < 10^{-10}$) down to below 20 mK in an
attempt to test the stress-strain relation for extremely small perturbations
and at very low temperature.

The experimental cell consists of a brass cylinder with internal diameter
$\diam$=15 mm, height $h$= 20.7 mm, and 0.3 mm thick walls, immersed in the main
body of the sample container, which can be cooled to below 20 mK by a
$^3$He-$^4$He dilution refrigerator. A broadband piezoelectric transducer,
made of 12 layers of a PVDF film folded on itself and located at the bottom of
the cell, can induce a displacement of the solid helium at frequencies from
the sub-audio range to above 20 kHz. The resulting strain field can be
detected at the top of the resonator by an ultra-sensitive displacement
sensor. The sensor consists of a flexible Kapton membrane with a
superconducting coating, the position of which is read by an electrodynamic
circuit with a dc-SQUID amplifier to a resolution in excess of $10^{-15}$
metre.

\begin{figure}[b]
\includegraphics[width=75 mm]{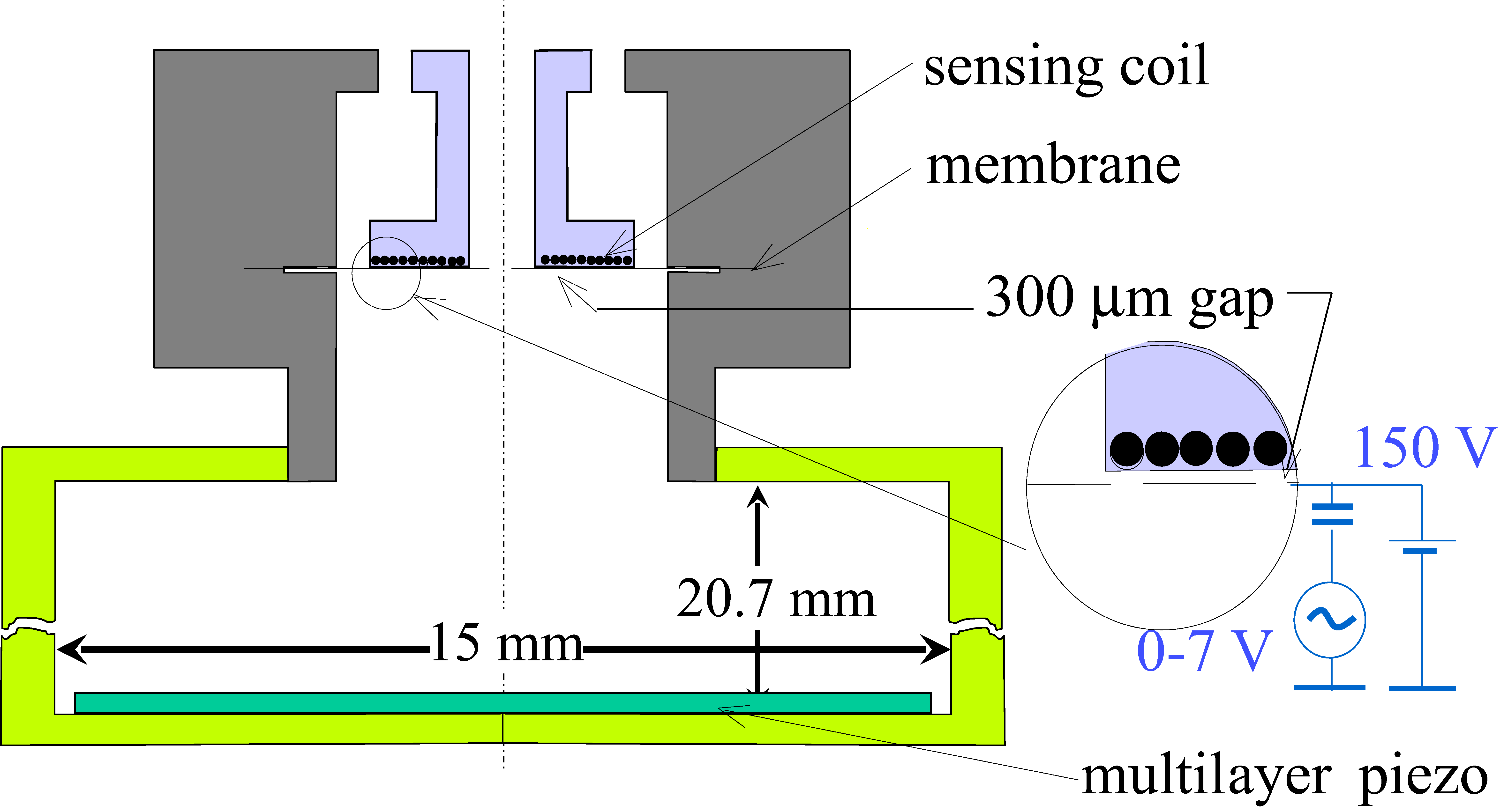}
\caption{\label{Cell}Schematic drawing of the acoustic cell, The displacement
  gauge is located at the top and the piezoelectric transducer at the
  bottom. The drawing is approximately to scale, except for the cut that makes
  it appear shrunk vertically.
} 
\end{figure}

The gap between the flexible diaphragm and the backplate in which the
superconducting sensing coil is embedded, is quite narrow ($\sim 300$
{\mbox{\small\greeksym{m}}}m). If a pressure $P_\m{el}$ is applied by the
diaphragm (here an electrostatic force) to the solid in the gap, the response
is expressed by the following stress-strain relation \cite{Landau:59}:
\begin{equation}        \label{uniaxial}
%  \sigma_{zz} = P_{\m{el}} = 3K\frac{1-\nu}{1+\nu} \epsilon_{\m zz}
  \sigma_{zz} = P_{\m{el}} = 3K[(1-\nu)/(1+\nu)]\, \epsilon_{\m zz}
             = M \epsilon_{\m zz}  \; .
\end{equation}
Equation (\ref{uniaxial}), where $K$ is the bulk modulus, $\nu$ Poisson's
ratio, and $M$ the uniaxial compression modulus, strictly holds
for isotropic materials (which hcp $^4$He is not).  We have no control over
the actual crystal orientation although we have reasons to believe that, most
often, the c-axis lies in a horizontal plane.  The strain $\epsilon_{zz}$ in
the flat gap of thickness $t$ equals the measured displacement $u_z$ of the
diaphragm divided by $t$. The strain arising from the side open to the main
volume of the cavity is reduced with respect to the thin side by approximately
the ratio $t/\diam$ of the gap thickness to the cavity diameter.

The piezoelectric transducer, when excited by an ac-voltage, generates in the
solid a nearly planar displacement along the vertical axis. Away from the
cavity acoustic frequencies at which inertia becomes important, and neglecting
bulk forces such as arising from gravitation or thermal gradients, the
equilibrium displacement field in the cylinder depends only on the
displacement at the boundaries and Poisson's ratio $\nu$ \cite{Landau:59a}.

When the drive frequency is increased from the low audio to the kHz range,
inertia becomes important and acoustic waves, longitudinal and transverse, can
propagate through the sample \cite{CommentQFS06}.  Standing waves are then
sustained when the acoustic half-wavelength matches the size of the
cavity, giving rise to acoustic resonances.

\begin{figure}[t]
\includegraphics[width=80 mm]{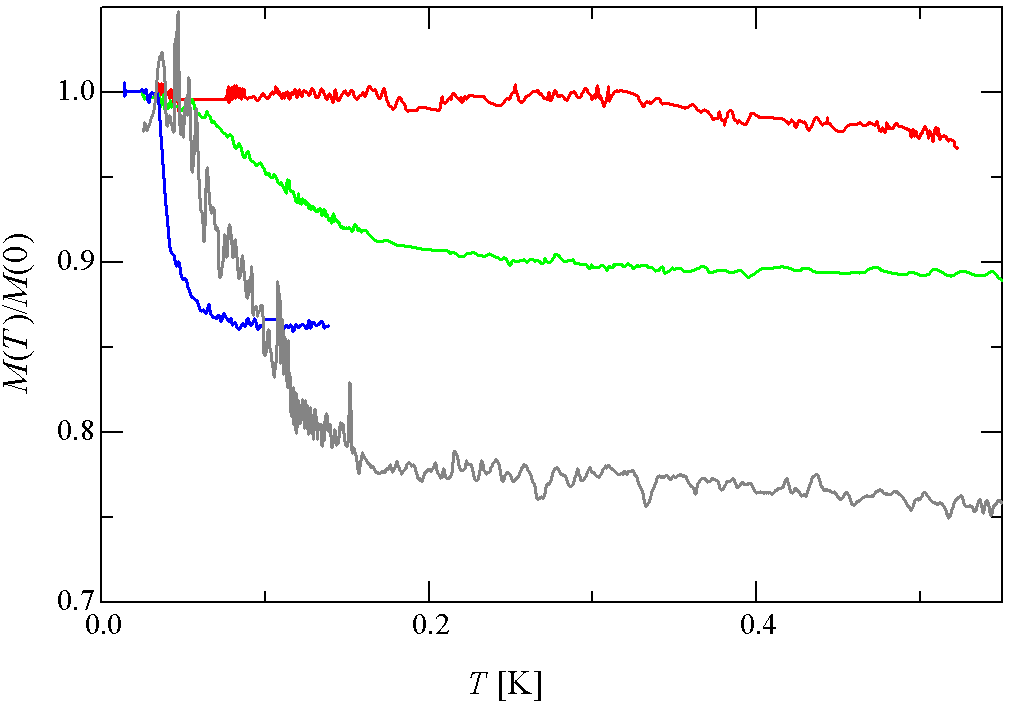}
\caption{\label{Compression} Reduced uniaxial compression modulus
  $M(T)/M(0.03)$) in terms of the temperature, in Kelvin.  The curves that
  extend to above 0.55 K pertain to a typical BC sample at a molar volume of
  20.65 cm$^3$ subjected to different thermal histories (bottom curve, fast
  cooling, top curve, slow cooling followed by rapid warming to 1.05 K), the
  two remaining curves to two CT samples, grown at 0.8 K (top) and 1.1 K
  (truncated at 0.15 K), slowly cooled to low temperatures. }
\end{figure}

The eigenfrequencies of the axisymmetric modes of a solid elastic cylinder
with rigid boundary conditions, which models the acoustic cavity used here,
have been studied by Hutchinson \cite{Hutchinson:66} among others. The
resonances in the finite cylinder closely track those obtained for waves
propagating in the infinite cylinder with displacement clamped at the walls
and restricted by periodic boundary conditions (zero displacement) at the
location of the cavity top and bottom ends. The resonance frequencies for the
various modes, normalized to $\omega_0= 2\,c_{\perp}/\diam$, $c_{\perp} =
[3K(1-2\nu)/2\rho(1+\nu)]^{1/2}$ being the transverse sound velocity
(isotropic case), can be expressed in terms of the cylinder aspect ratio
$h/\diam$ and Poisson's ratio $\nu$. The fundamental mode reduced frequency is
equal to $\omega/\omega_0=\alpha(h/\diam,\nu)=3.07$ for $h/ \diam=1.38$ and
$\nu \sim 0.3$ ~\cite{Poisson} with only a weak dependence on $\nu$ around
$\nu$=0.3 ~\cite{Hutchinson:66}. A direct numerical simulation using the exact
cavity geometry yields the same figure to better than 1 \%.

\begin{figure}[tb]
 \begin{center}
  \hskip -8mm  
  \includegraphics[width=80mm]{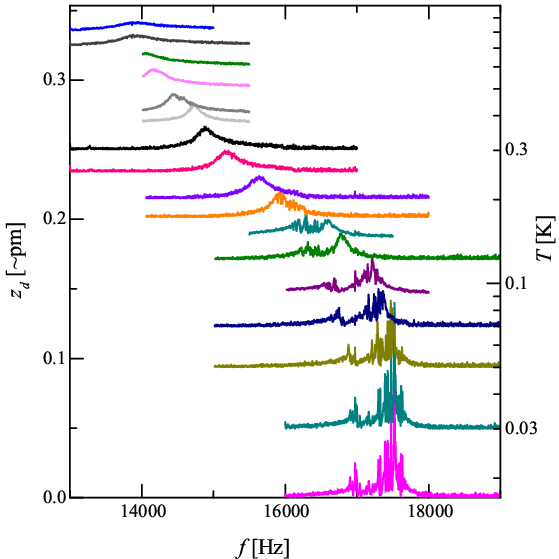}  
  \caption{\label{Spectra}Acoustic spectra at various temperatures for a
    QC sample with molar volume 20.65 cm$^3$; left $y$-axis, relative diaphragm
    displacement $z_\m d$ in picometres; right $y$-axis, temperature in Kelvin
    on a logarithmic scale; $x$-axis, frequency in Hertz. Below 200 mK, the
    resonance line splits and develops wiggles; a satellite also develops
    to the left of the main peak.}
 \end{center}
\end{figure}

Thus, the acoustic cell can be operated in three different ways: (1) the
variation of uniaxial compression in the displacement sensor gap; (2) that of
the transverse velocity (that is, of shear modulus) in the acoustic cavity by
the resonance frequency of the fundamental mode; (3) a direct, albeit less
precise, check on the variation of $\nu$ in the cavity by the diaphragm
response to a quasi-static (70 Hz) actuation of the piezo. Typical results for
measurements (1) and (2) are shown on Fig.\ref{Compression} and \ref{Spectra}
for several samples.  These results depend, apart from the cavity geometry, on
two quantities, the bulk modulus divided by the density $K/\rho$ and
Poisson's ratio $\nu$. Since no pressure variation is recorded when the sample
is cooled at constant volume the bulk modulus $K$ remains constant. Its value
is known from the phase diagram and from sound velocity measurements
\cite{Greywall:71}.

The solid $^4$He samples have been obtained by the two conventional methods.
With the blocked capillary (BC) method, helium is solidified by filling the
cell to a given density (or a given pressure, usually 56 bar here), cooled
through complete freezing, and then further down in temperature, the
quantity of helium in the cell remaining constant because the solid formed in
the filling capillary makes an immovable plug at low temperature. This method
is used to make samples with densities higher than the density close to
melting but subsequent cooling induces thermal stresses that usually damage
the crystal. To obtain crystals with lower internal stress, the solid can be
grown at low temperature by feeding liquid into the cell at constant
temperature (CT), producing samples with a density close to that at melting.
Solid $^4$He is mechanically fragile: defects can be created by mild
mechanical shocks and high intensity sound bursts. The samples were grown from
industrial grade helium gas, which contains between 0.1 (as measured in
recycled gas) to 0.3 ppm (nominal from the well tap) of $^3$He impurities.
Upon thermal cycling, crystals may split into crystallites with different
orientations, break and regrow, the dislocation line density may either
decrease or increase, the $^3$He impurities diffuse about onto lower density
regions - grain boundaries, liquid inclusions, and, quite importantly here, on
dislocation cores, providing pinning sites.

The temperature variation of the uniaxial compression modulus $M$, given by
the diaphragm displacement under a given applied force, is shown in
Fig.\ref{Compression} for a BC sample of molar volume 20.65 cm$^3$, determined
by an ancillary quartz resonator in the liquid phase, and different cooling
rates, as well as two CT samples grown at different temperatures (0.8 and 1.1
K). The CT ($T=1.1$ K) sample shows a marked drop below 50 mK, the other CT
($T=0.8$ K) sample shows no variation below $\sim 0.3$~K down to 33 mK. Most
samples however do show a stiffening at low temperatures that varies in
overall magnitude and onset temperature according to their thermal and
mechanical history. These observations fully vindicate previous studies
\cite{Wanner:76,Iwasa:79,Tsuruoka:79,Paalanen:81,Hiki:83,Day:09}, including
the markedly different behaviors of CT samples \cite{Paalanen:81}.

When the drive frequency is scanned, using the piezo as a transmitter and the
diaphragm as a microphone, the fundamental resonance frequency yields, in a
first approximation as discussed above, the change in the shear modulus
$G_{\m{eff}}=3K(1-2\nu)/2(1+\nu)=\rho\,c_\perp^2$ with temperature. An example
of such scan is given in Fig.\ref{Spectra}; the frequency of the resonance
peak decreases by 20 \% from 17 to 717 mK, which implies that the effective
shear modulus $G_{\m{eff}}$ of this particular sample is divided by 1.58 upon
warming.

\begin{figure}[t]
 \includegraphics[width=80 mm]{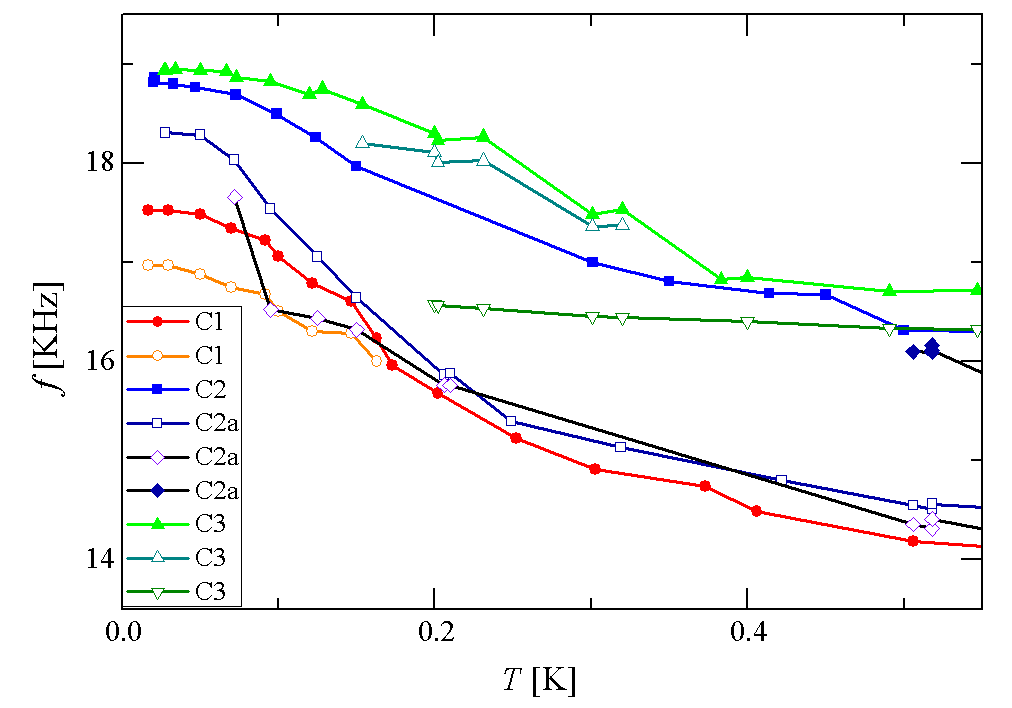}  
 \caption{\label{FrequencyTemperature}Acoustic resonance peak frequency vs
   temperature for several BC samples: C1, quenched, same as in
   Fig.\ref{Spectra}; C2, 20.87 cm$^3$ per mole, slowly cooled, C2a, after
   thermal shock; C3, 20.33 cm$^3$ per mole.  }
\end{figure}

% The table with data for Cr1-Cr3 is here: 
% T  Vmol P Sample
% 2.426 20.69643 56.43213 Cr1
% 2.428 20.87182 55.21983 Cr2
% 2.527 20.3255 59.49537 Cr3

The resonance frequencies in different samples are shown in terms of the
temperature in Fig.\ref{FrequencyTemperature}. As for the uniaxial compression
modulus, the frequency shifts are markedly sample dependent. The variations in
the low temperature limit for samples grown at the same density reflect
variations in the crystal orientation and (or) its splitting into several
grains, or also the presence of residual stress and pressure
inhomogeneities. It is noteworthy that the CT sample that showed no variation
in $M$ in Fig.\ref{Compression} exhibits no frequency shift either in
Fig.\ref{FrequencyTemperature}. Uniaxial compression applies shear in two
different planes, at least one of which is not parallel to the glide plane of
the dislocations. The acoustic resonance generates strains in all
directions. Dislocations, if present and unpinned, should have contributed to
the total strain, which they did not.

In most samples, two different regimes appear below and above the temperature
$T^*$ around which the frequency shifts at the fastest rate ($\sim$ 200 mK in
Fig.\ref{Spectra}), which is also the temperature at which the resonance
amplitude goes through a minimum. Contrarily to other work \cite{Day:09}, we
do not observe a narrowing of the resonance line, but a splitting into several
resonances, or wiggles, below $T^*$. This splitting is more pronounced at the
lowest temperature, where the crystal is stiffest and intrinsic damping
lowest. On occasions, distinct, weak, resonance peaks also show up several
hundreds of Hz above or below the main resonance line. Both features, which
can be seen in Fig.\ref{Spectra}, appear in the linear response regime. Their
origin is unknown but they reveal that the resonance line is inhomogeneously
broadened at $T\gtrsim T^*$ and partially resolved below, presumably because
damping decreases \cite{artefact}.  The fundamental mode depends on some
coarse-grained {\it{average}} of the density and elastic properties of the
medium in the acoustic cavity and does not probe inhomogeneities or localized
defects. We believe that the splitting and the satellite lines arise from the
existence of additional, macroscopic, degrees of freedom within the solid
sample as, for instance, a condensate fraction, or a dislocation network
collective motion.

At higher drive levels, non-linear effects appear, notably frequency pulling
that tends to push the resonance frequency back to its high temperature value.
The drive amplitude at which this effect sets in depends on temperature; it
goes through a minimum around 100 to 200 mK. The value of the
strain is $\simeq 10^{-9}$, an order of magnitude lower than the threshold
reported in Ref.[\onlinecite{Syshchenko:09}]. It should be noted that the
corresponding displacement at its maximum, at the center of the cavity, is
exceedingly small, less than one tenth of the width of the Peierls valley.

From the low $T$ limit of the resonance frequency in Fig.\ref{Spectra},
$\omega/2\pi=17.5$ kHz, and $\alpha = 3.07$, we find, using for the value of
the bulk compression modulus at a molar volume of 20.65 cm$^3$, $K=275$ bar
\cite{BulkModulus}, a value for Poisson's ratio of 0.28 at 0.02~K, not much
different from the value obtained by high frequency sound measurements
\cite{Poisson}, climbing back to 0.36 at 0.7 K with the appropriate value of
$\alpha$ (3.20). A similar variation is obtained from the change in uniaxial
compression modulus in Fig.\ref{Compression}.

Most of the features reported here find a natural explanation in terms of
dislocation motion in the framework of the Granato-L\"ucke theory as used by
many authors over the years (see, e.g.
\cite{Wanner:76,Tsymbalenko:77,Iwasa:79,Tsuruoka:79,Paalanen:81,Day:09,%
  Syshchenko:09} and references therein) and may also be related
to the anomalies in the TO experiments as discussed in
Refs.\cite{Clark:08,Day:09}.  Other features do not. The behavior of the CT
samples, showing a more abrupt (or else, very little) change in $G_{\m{eff}}$
at lower $T$ requires a like-sudden change (or lack of) in dislocation
pinning. Such behaviors have also been observed in TO experiments on the NCRI
fraction in good quality samples \cite{Clark:07}.  Sometimes, the temperature
dependence of $G_{\m{eff}}$ shows a kink, as can be seen in
Fig.\ref{Compression} for the fast-cooled BC sample. This is not really
expected in the dislocation line model, which relies on the condensation of
$^3$He impurities on dislocation cores and should lead to a more gradual
variation, like for the curve obtained after warming the sample to 1.05 K.

Let us nonetheless interpret the drastic change in elastic properties observed
here in terms of the strain induced by dislocations, following Paalanen et al.
\cite{Paalanen:81} and using their expression
\begin{equation}        \label{EffectiveShear}
  G/G_{\m{eff}} = 24 R (1-\nu) \Lambda L^2/\pi^3\,+\,1 \; ,
\end{equation}
$R$ depending on the orientation between the local displacement vector and the
$c$-axis of the crystal. Assuming equal weight between longitudinal and
transverse motions, $R \leqslant 1/4$ and using our result
$G/G_{\m{eff}}=1.58$ and $\nu=0.3$, we find $\Lambda L^2 \geqslant 4.2$.  This
value greatly exceeds that found experimentally by Iwasa et al.
\cite{Iwasa:79} (0.21) and justified theoretically by them for the hcp
lattice, and is also larger than that quoted by Paalanen et al. ($\simeq 2.0$)
\cite{Paalanen:81}.  The stiffness of hcp $^4$He decreases very markedly above
$\sim 0.1-0.2$ K in impure, quenched BC crystals, above $\sim 50-70$ mK in
better quality CT crystals. The corresponding value of $\Lambda L^2$ is
anomalously large, indicating that the dislocation network is highly
disorganized. Dislocations seem to proliferate and creep over all the space
available to them even for extremely small strains as if Peierls' barrier was
irrelevant.

We thank S\'ebatien Balibar, Giulio Biroli, Jean-Philippe Bouchaud, and
Francesco Zamponi for stimulating discussions from which we have greatly
benefitted.

% \bibliography{/home/vx/texjob/Supersolid/PRL08/PRL08}

\begin{thebibliography}{23}
\bibitem[{art()}]{artefact}
\bibinfo{note}{It is unlikely that the wiggles would have arisen from some
  unsuspected experimental artefact such as a parasitic resonance. The wiggles
  are not seen in liquid where lines are sharp. They come in different forms
  with different solid samples. All samples except the CT sample grown at 0.8 K
  show a split fundamental resonance.}
\bibitem[{Bul()}]{BulkModulus}
\bibinfo{note}{From Fig.17 in Ref.\cite{Greywall:71}. The density in the cavity
  can be less than that in the main cell for quenched crystals.}
\end{thebibliography}

\end{document}